# UVG-VPC: Voxelized Point Cloud Dataset for Visual Volumetric Video-based Coding

Guillaume Gautier, Alexandre Mercat, Louis Fréneau, Mikko Pitkänen, and Jarno Vanne
Ultra Video Group, Tampere University, Tampere, Finland
{guillaume.gautier, alexandre.mercat, louis.freneau, mikko.pitkanen, jarno.vanne}@tuni.fi

*Abstract*— Point cloud compression has become a crucial factor in immersive visual media processing and streaming. This paper presents a new open dataset called UVG-VPC for the development, evaluation, and validation of MPEG Visual Volumetric Video-based Coding (V3C) technology. The dataset is distributed under its own non-commercial license. It consists of 12 point cloud test video sequences of diverse characteristics with respect to the motion, RGB texture, 3D geometry, and surface occlusion of the points. Each sequence is 10 seconds long and comprises 250 frames captured at 25 frames per second. The sequences are voxelized with a geometry precision of 9 to 12 bits, and the voxel color attributes are represented as 8-bit RGB values. The dataset also includes associated normals that make it more suitable for evaluating point cloud compression solutions. The main objective of releasing the UVG-VPC dataset is to foster the development of V3C technologies and thereby shape the future in this field.

## I. Introduction

Recent advances in volumetric visual media technologies have opened a plethora of opportunities for *Extended Reality* (*XR*). The state-of-the-art volumetric sensing and capturing technologies allow for the creation of detailed and immersive digital representations of the real world in *three-dimensional* (*3D*) space. In general, these representations can be represented as polygon meshes or point clouds that provide a realistic and detailed view of scenes from any viewpoint. Moreover, the natural and realistic viewing experience in XR is enhanced by *6 degrees of freedom* (*6DoF*), which enables viewers to move around in the scene with both translational and rotational freedom and thereby expand the viewing space.

Economic storage and transmission of volumetric visual data require efficient compression technologies. To that end, the *Motion Picture Experts Group* (*MPEG*) has released the *Visual Volumetric Video-based Coding* (*V3C*) standards ISO/IEC 23090-5 [1] to compress dynamic volumetric scenes for XR applications, including gaming, sports broadcasting, and motion pictures. V3C can be used to compress various types of volumetric content, such as point clouds, immersive video with depth, and mesh representations of visual volumetric frames. For the time being, V3C includes two standards: *Video-based Point Cloud Compression* (*V-PCC*) [2], and *MPEG immersive video* (*MIV*) [3], [4], of which this paper focuses on the V-PCC standard.

*Rate-distortion* (*RD*) performance of video codecs is typically evaluated with objective and subjective metrics, which involves a trade-off between coding efficiency and loss of information. Conducting these quality assessments comprehensively calls for representative datasets that cover a broad range of content (e.g., motion, texture, or occlusion).

This work was supported in part by the XR Simulation and Presence at the Cloud Edge (XR-SPACE) project led by Nokia and funded by Business Finland, and the Academy of Finland (decision no. 349216).

Given that V-PCC is commonly used for telecommunication and XR applications [4], the test set should be composed of voxelized point cloud full-body human subjects, as shown by the *Common Test Conditions* (*CTC*) for V-PCC [5].

Table I lists the existing open point cloud datasets of full human body. The first two of them are not voxelized [6], [7], whereas the remaining three have limitations in geometry precision and size [8]–[10]. Hence, none of them is optimal for the development and evaluation of V-PCC tools, so there is an urgent need for high-quality datasets that contain real-world scenes with multifaceted content and motion.

This paper presents a new open dataset called UVG-VPC that is made up of 12 voxelized point cloud test sequences. Each sequence is 10 s in length, comprises 250 frames captured at a frame rate of 25 *frame per second* (*fps*), and has RGB attribute precision of 8 bits and a geometry precision of 9, 10, 11, and 12 bits. The dataset is available online at

https://ultravideo.fi/UVG-VPC/

It is released under its own non-commercial license [11]. Additionally, the associated normals are provided for all sequences. To the best of our knowledge, the proposed dataset is the first and only one that has been entirely designed for the development, evaluation, and validation of V-PCC coding technologies. The UVG-VPC dataset seeks to serve as a valuable resource for researchers and practitioners in the field of volumetric data compression and beyond.

The remainder of the paper is outlined as follows. Section II describes the volumetric capture studio setup used to obtain the needed data. Section III details the proposed workflow for voxelized point cloud generation. Section IV introduces our UVG-VPC dataset and discusses its characteristics. Finally, Section V concludes the paper.

## II. Volumetric Capture Studio

The proposed dataset was captured with the volumetric capture studio developed by *Mantis Vision* [12]. Fig. 1 illustrates the studio setup that is composed of 32 (19 *long* and 13 *short*) camera units with different stereo distances. Each camera unit is composed of two RGB cameras, an IR projector, an *IR (infrared)* camera, and an *Intel Next Unit of*

TABLE I. Existing Open Point Cloud Datasets of Full Human Body

| Ref | Dataset | #seq. | Fps | #frames | #cams | Voxelized/ geometry precision |
|---|---|---|---|---|---|---|
| [6] | CWIPC-SXR | 21 | 30 | 596–2768 | 7 | No |
| [7] | Volograms & V-SENSE | 3 | 30 | 149–1830 | 12/60 | No |
| [8] | 8iVFBv2* | 4 | 30 | 300 | 42 | Yes/10bits |
| [9] | Owlii* | 4 | 30 | 600 | - | Yes/11bits |
| [10] | 8iVSLF | 7 | 30 | 1,300 | 39 | Yes/12bits |
| **Our** | **UVG-VPC** | **12** | **25** | **250** | **96** | **Yes/9–12bits** |

*Partially included in the CTC for V-PCC [5].

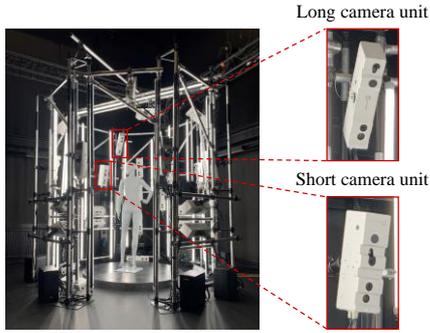

Fig. 1. Volumetric capture setup.

TABLE II. SPECIFICATION OF SHORT AND LONG CAMERA UNITS

| Camera unit type | | Long (×19) | Short (×13) |
|---|---|---|---|
| **RGB camera ×2** | Specification | UI-328xCP-C | UI-308xCP-C |
| | Resolution | 2456×2054 | 2456×2054 |
| | Stereo distance | ~30cm | ~10cm |
| **IR camera** | Specification | UI-314xCP-M | |
| | Resolution | 640×512 | |
| **Intel NUC** | Processor | Intel(R) Core(TM) i7-8665U CPU @ 1.90GHz | |
| | Memory | 32 GB | |
| | Hard drive | Samsung 970 EVO Plus SSD 1Tb | |

*Computing* (*NUC*); their specifications are given in Table II. The camera units are connected through a tree topology using 10Gbps switches, with four camera units connected to each switch, and two switches connected to the render computer. The studio is able to capture volumetric video at up to 25 fps.

The studio features 40 LED tubes of 50 W and a Sync LED that flashes and triggers at the recording frame rate. Cameras capture images at slightly different times to avoid IR interference, with opposite cameras exposing at the same time. The studio is set up to a height of 2.5 m, with a diameter of 3 m, and it allows scanning of a scene with a height of 2.2 m and a diameter of 1.6 m. To enhance the capture quality of faces, most cameras are located on the upper part of the body.

### III. VOXELIZED POINT CLOUD GENERATION

Fig. 2 depicts the five steps needed to generate the proposed voxelized point cloud dataset. The first two steps are processed by the off-the-shelf equipment of the volumetric capture studio. The remaining three steps are designed in this work to make our sequences matching the format of the sequences from CTC for V-PCC [5].

1) **Point cloud acquisition** is executed by the camera units and the render computer. The camera units capture both RGB and IR data, which are fused by the NUCs into a point cloud structure. Both camera data and generated point clouds are sent over the network to the render computer that merges them into a single *Raw Merged Point Cloud*.

2) **Mesh generation** is used to create a *Mesh* from the *Raw Merged Point Cloud* with off-the-shelf Poisson surface reconstruction algorithm provided with the volumetric capture studio.

3) **Mesh sampling** deploys triangle point picking from the *trimesh* Python library [13] to generate a *Sampled Mesh Point Cloud*. After studying the trade-off between having a sufficient number of points for the following voxelization process and a reasonable memory footprint, the number of sampled points was fixed to 10 million.

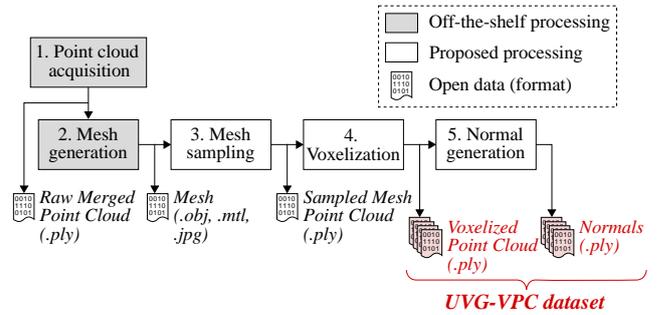

Fig. 2. Overview of the UVG-VPC dataset generation process.

4) **Voxelization** is the process of applying point cloud data on a regular 3D grid structure, where each cell or voxel represents the presence or absence of points within its boundaries. When multiple points are involved, color attributes are averaged. For this step, the voxel size is computed as the maximum dimension of capture system bounding box across all dimensions divided by $2^N$, where $N$ is the geometry precision. The UVG-VPC dataset includes *Voxelized Point Clouds* with a geometry precision of 9, 10, 11, and 12 bits.

5) **Normal generation** computes the *Normals* for each *Voxelized Point Cloud* in the UVG-VPC dataset using *open3D* Python library [14] and a Knn normal estimation with 12 neighbours [15], [16]. These *Normals* are used in the CTC for V-PCC to calculate the quality metric known as D2 [5]. Providing the *Normals* enables fair comparisons between solutions.

In addition to the UVG-VPC dataset, we provide open access to the intermediate data used to create the sequences. Scientific community is free to use it in voxelized point cloud generation with varying geometry precision, as well as in other areas of interest such as mesh generation or dynamic mesh compression [17].

### IV. UVG-VPC DATASET

The proposed UVG-VPC dataset consists of 12 sequences, each 10 s long and composed of 250 frames captured at 25 fps. For each sequence, point cloud voxelized at 9, 10, 11, and 12 bits are provided with their associated normals.

Table III lists the UVG-VPC sequences alphabetically and characterizes them with snapshots, names, content descriptions, and specific features. There is also a graph for each sequence that shows the distribution of points per frame for a geometry precision of 10 bits, as well as the corresponding average, minimum, and maximum values.

The features of the sequences were carefully selected to make them challenging for various compression algorithms and ensure that the dataset is representative of real-world scenarios. In particular, the characterisation was done with respect to the following features:

- speed: speed of moving points;
- motion field: quantity of moving points;
- RGB texture: texture complexity of the RGB attributes;
- 3D geometry: complexity of the volumetric shapes; and
- surface occlusion: number of (dis)appearing points.

Table IV summarizes the characteristics of the UVG-VPC sequences. They all have unique characteristics, i.e., no two sequences sharing the same set of features. Furthermore, some sequences were specifically designed to contrast with each other in terms of one or more of these criteria.

TABLE III. CHARACTERISTICS OF THE VOXELIZED POINT CLOUD SEQUENCES IN THE PROPOSED UVG-VPC DATASET

| Snapshot | Name and description | #pts/frame (10-bits) | Snapshot | Name and description | #pts/frame (10-bits) |
|---|---|---|---|---|---|
| 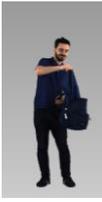 | **Name:** BlueBackpack **Description:** a person takes a jacket out of a backpack and puts it on. **Specific features:** unicolor clothes and accessories; the interaction with the accessories adds 3D geometry complexity and surface occlusions. | 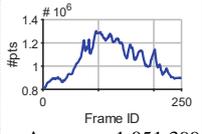 **Average:** 1 051 399 **Min:** 799 322 **Max:** 1 302 904 | 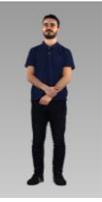 | **Name:** BlueSpin **Description:** a person dressed in blue is steadily spinning around. **Specific features:** unicolor clothes; steady rotation about a fixed axis. | 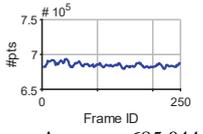 **Average:** 685 044 **Min:** 679 347 **Max:** 693 684 |
| 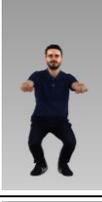 | **Name:** BlueSquat **Description:** a person dressed in blue is performing squats. **Specific features:** unicolor clothes; intermittent surface occlusions due to body movements. | 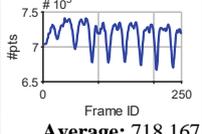 **Average:** 718 167 **Min:** 667 024 **Max:** 741 177 | 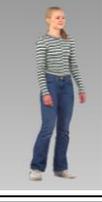 | **Name:** CasualSpin **Description:** a person wearing a striped shirt and jeans is steadily spinning around. **Specific features:** textured top, unicolor bottom; steady rotation about a fixed axis. | 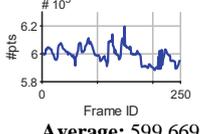 **Average:** 599 669 **Min:** 588 394 **Max:** 619 662 |
| 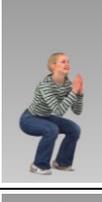 | **Name:** CasualSquat **Description:** a person wearing a striped shirt and jeans is performing squats. **Specific features:** textured top; unicolor bottom; intermittent surface occlusions due to body movements. | 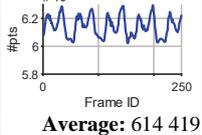 **Average:** 614 419 **Min:** 602 150 **Max:** 629 416 | 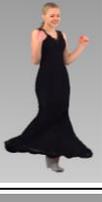 | **Name:** ElegantDance **Description:** a person wearing a long black dress is dancing and twirling around. **Specific features:** unicolor clothes; as the person moves and twirls, the dress is folding and shifting, creating a complex 3D geometry. | 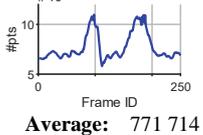 **Average:** 771 714 **Min:** 579 917 **Max:** 1 098 016 |
| 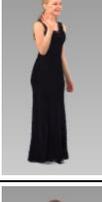 | **Name:** ElegantWave **Description:** a person wearing a long black dress greets by waving hand. **Specific features:** unicolor clothes; only upper-body movement; no dress movement; simple 3D geometry. | 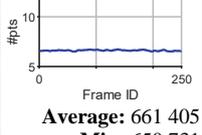 **Average:** 661 405 **Min:** 650 731 **Max:** 674 383 | 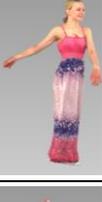 | **Name:** FlowerDance **Description:** a person wearing a long flower dress is dancing and twirling around. **Specific features:** textured clothes; as the person moves and twirls, the dress is folding and shifting, creating a complex 3D geometry. | 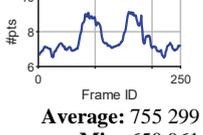 **Average:** 755 299 **Min:** 650 961 **Max:** 913 621 |
| 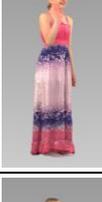 | **Name:** FlowerWave **Description:** a person wearing a long flower dress greets by waving hand. **Specific features:** textured clothes; only upper-body movement; no dress movement; simple 3D geometry. | 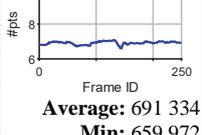 **Average:** 691 334 **Min:** 659 972 **Max:** 708 898 | 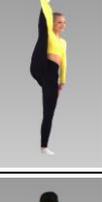 | **Name:** Gymnast **Description:** a person stands on one leg and does a leg hold. **Specific features:** unicolor clothes; leg movement creates surface occlusion. | 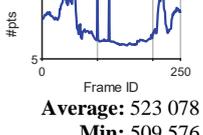 **Average:** 523 078 **Min:** 509 576 **Max:** 551 604 |
| 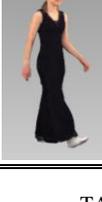 | **Name:** HelloGoodbye **Description:** a person wearing a long black dress enters the scene, greets by waving hand, and leaves the scene. **Specific features:** unicolor clothes; empty capture space at sequence start and end. | 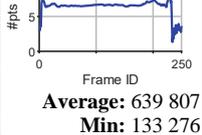 **Average:** 639 807 **Min:** 133 276 **Max:** 929 588 | 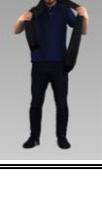 | **Name:** ReadyForWinter **Description:** a person puts on a beanie and a scarf. **Specific features:** unicolor clothes; textured scarf; interaction with accessories creates complex surface structures and surface occlusions. | 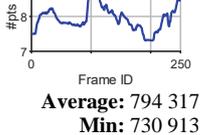 **Average:** 794 317 **Min:** 730 913 **Max:** 883 642 |

TABLE IV. UVG-VPC DATASET CHARACTERIZATION

| Sequence | Speed | Motion field | RGB texture | 3D geometry | Surface occlusion |
|---|---|---|---|---|---|
| BlueBackpack | Fast | Medium | Simple | Complex | Plenty of |
| BlueSpin | Medium | Dense | Simple | Simple | Little |
| BlueSquat | Fast | Dense | Simple | Medium | Medium |
| CasualSpin | Medium | Dense | Medium | Simple | Little |
| CasualSquat | Fast | Dense | Medium | Medium | Medium |
| ElegantDance | Fast | Dense | Simple | Complex | Plenty of |
| ElegentWave | Slow | Sparse | Simple | Simple | Little |
| FlowerDance | Fast | Dense | Complex | Complex | Plenty of |
| FlowerWave | Slow | Sparse | Complex | Simple | Little |
| Gymnast | Medium | Medium | Simple | Simple | Medium |
| HelloGoodbye | Medium | Medium | Simple | Medium | Plenty of |
| ReadyForWinter | Medium | Medium | Medium | Complex | Plenty of |

## V. CONCLUSION

This paper presented the UVG-VPC open dataset, which has been carefully designed to facilitate the development, evaluation, and validation of V-PCC coding technology. The dataset consists of 12 voxelized point cloud sequences and associated normals. We believe that the availability of the UVG-VPC dataset will enable researchers and practitioners to advance the state-of-the-art in point cloud compression and foster its deployment in immersive visual media applications.

ACKNOWLEDGMENT

This work was carried out with the support of *Centre for Immersive Visual Technologies* (*CIVIT*) research infrastructure, Tampere University, Finland. In addition, the authors wish to acknowledge CSC – IT Center for Science, Finland, for computational and storage resources.

## REFERENCES

[1] ISO/IEC 23090-5:2021. "Information technology — coded representation of immersive media — part 5: visual volumetric video-based coding (V3C) and video-based point cloud compression (V-PCC)," Jun. 2021.

[2] D. Graziosi, *et al.*, "An overview of ongoing point cloud compression standardization activities: video-based (V-PCC) and geometry-based (G-PCC)," *APSIPA Trans. Signal Inf.ormation Process.,* vol. 9, Apr. 2020.

[3] J. M. Boyce, *et al.*, "MPEG immersive video coding standard," *Proc. IEEE,* vol. 109, no. 9, pp. 1521–1536, Sep. 2021.

[4] V. K. M. Vadakital, *et al.*, "The MPEG immersive video standard—current status and future outlook," *IEEE MultiMedia*, vol. 29, no. 3, pp. 101–111, Jul.–Sep. 2022.

[5] ISO/IEC JTC1/SC29/WG11, "Common test conditions for V3C and V-PCC," Document N19518, Online, Jul. 2020.

[6] I. Reimat, *et al.*, "CWIPC-SXR: point cloud dynamic human dataset for social XR," in *Proc. ACM Multimedia Sys. Conf.*, pp. 300–306, Istanbul, Turkey, Sep. 2021.

[7] R. Pagés, K. Amplianitis, J. Ondrej, E. Zerman, and A. Smolic, "Volograms & V-SENSE volumetric video dataset," Mar. 2022.

[8] E. d'Eon, B. Harrison, T. Myers, and P. A. Chou, "8i voxelized full bodies - a voxelized point cloud dataset," ISO/IEC JTC1/SC29 Joint WG11/WG1, Document WG11M40059/WG1M74006, Geneva, Switzerland, Jan. 2017.

[9] Y. Xu, Y. Lu, and Z. Wen, "Owlii dynamic human mesh sequence dataset," ISO/IEC JTC1/SC29/WG11 Document M41658, Macau, China, Oct. 2017.

[10] M. Krivokuća, P. A. Chou, and P. Savill, "8i voxelized surface light field (8iVSLF) dataset," ISO/IEC JTC1/SC29 WG11, Document M42914, Ljubljana, Slovenia, Jul. 2018.

[11] UVG-VPC Licence, [Online], Available: https://ultravideo.fi/UVG-VPC/licence.pdf, Accessed: May. 26, 2023.

[12] Mantis Vision Website, [Online], Available: https://mantis-vision.com/, Accessed: Apr. 26, 2023.

[13] E. W. Weisstein, "Triangle point picking," [Online], Available: https://mathworld.wolfram.com/TrianglePointPicking.html, Accessed: Apr. 26, 2023.

[14] Q.-Y. Zhou, J. Park, and V. Koltun, "Open3D: a modern library for 3D data processing," arXiv:1801.09847, Jan. 2018.

[15] D. Tian, H. Ochimizu, C. Feng, R. Cohen, and A. Vetro, "Evaluation metrics for point cloud compression," ISO/IEC JTC1/SC29/WG11, Document M39966, Geneva, Switzerland, Jan. 2017.

[16] D. Tian, H. Ochimizu, C. Feng, R. Cohen, and A. Vetro, "Updates and integration of evaluation metric software for PCC," ISO/IEC JTC1/SC29/WG11, Document M40522, Hobart, Australia, Apr. 2017.

[17] M. Wien, J. Jung, and V. Baroncini, "Formal visual evaluation and study of objective metrics for MPEG dynamic mesh coding," in *Proc. Eur. Workshop Vis. Inf. Process.*, Lisbon, Portugal, Sep. 2022.